\documentclass[aip,reprint,showpacs,twocolumn,superscriptaddress,floatfix]{revtex4-1}
\usepackage{epsfig}
\usepackage{amsmath}
\usepackage{amssymb}
\usepackage{amsfonts}
\usepackage{mathptmx}
\usepackage{dcolumn}
\usepackage{eucal}
\usepackage{bm}
\usepackage{color}
\usepackage[colorlinks,linkcolor=blue,citecolor=blue]{hyperref}

\usepackage{epstopdf}

\usepackage{graphicx}

\usepackage{natbib}

\begin{document}
\title{The Superconducting Quantum Interference Proximity Transistor (SQUIPT)}
\author{F. Giazotto}
\email{giazotto@sns.it}
\affiliation{NEST, Instituto Nanoscienze-CNR and Scuola Normale Superiore, I-56127 Pisa, Italy}

%

\maketitle
\section{Introduction}
\label{intro}

The superconducting quantum interference device (SQUID) is recognized as the most sensitive magnetic-flux detector ever realized, and combines the physical phenomena of Josephson effect and flux quantization to operate. SQUIDs are nowadays exploited in a variety of physical measurements with applications spanning, for instance, from pure science to medicine and biology. Recently, the interest in the development of nanoscale SQUIDs has been motivated by the opportunity to exploit these sensors for the investigation of the magnetic properties of isolated dipoles with the ultimate goal to detect one single atomic spin, i.e., one Bohr magneton. 

Here we describe a hybrid superconducting interferometer which exploits the phase dependence of the density of states (DOS) (i.e., the physical quantity that refers to the number of states per unit energy at each energy and per volume available to be occupied by electrons) of a metallic nanowire placed in good electric contact with a superconductor to achieve high sensitivity to magnetic flux. The operation of a prototype structure based on this principle, the superconducting quantum interference proximity transistor (SQUIPT), has been recently reported. Limited power dissipation joined with the opportunity to access single-spin detection make this interferometer attractive for the investigation of the switching dynamics of individual magnetic nanoparticles. The device description is organized as follows. 
Proximity effect as well as the model of the hybrid superconducting magnetometer is presented in Sec. \ref{PE}. The device response in terms of voltage modulation and transfer function is shown in Sec. \ref{devres}. The noise behavior is presented in section \ref{noise} where the feasibility of this structure as a single-spin detector is also briefly addressed. Section \ref{real} is devoted to the presentation of the response of a real SQUIPT device.

\begin{figure}[t!]
\includegraphics[width=\columnwidth]{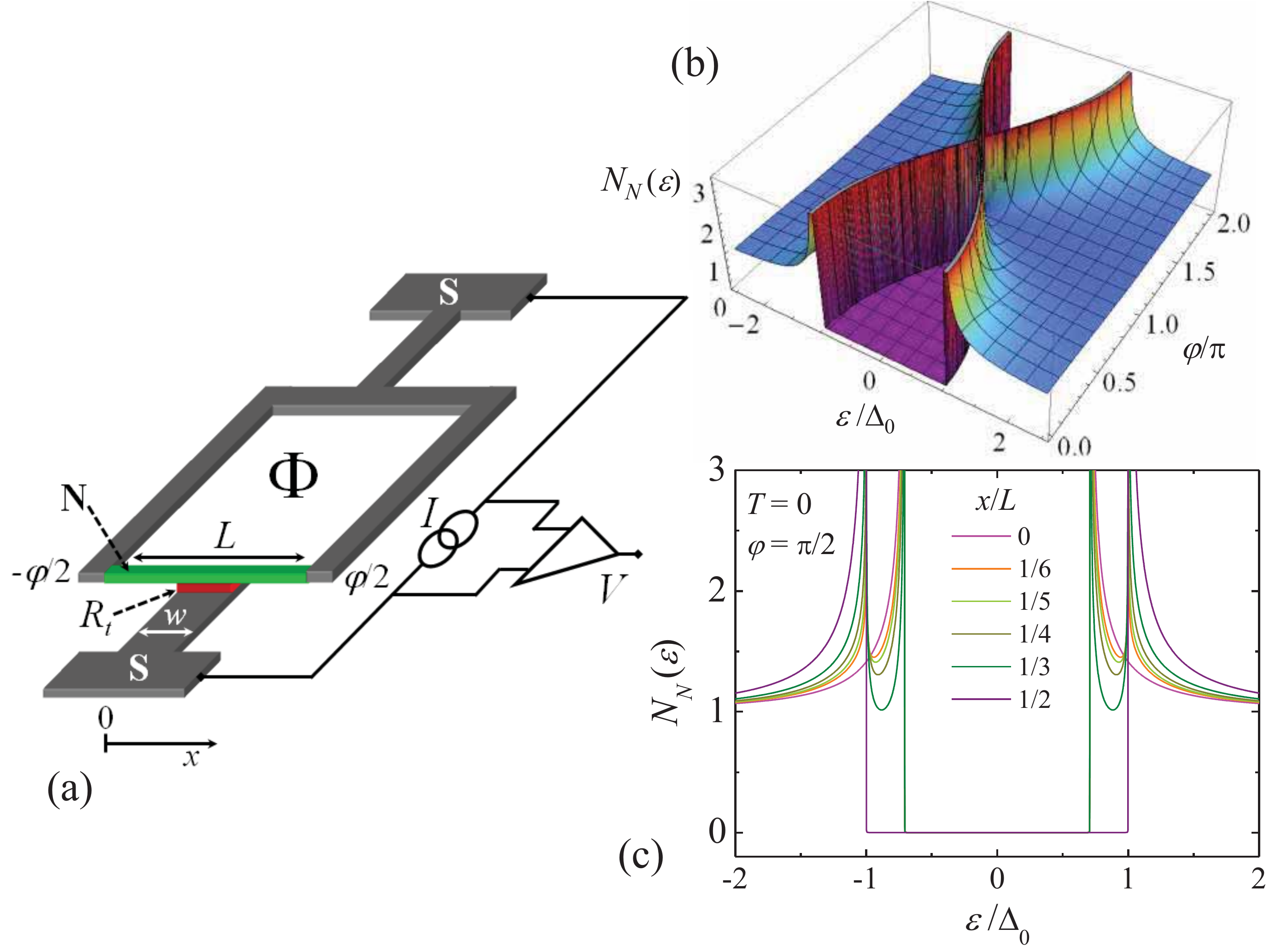}
\caption{\textbf{\emph{The SQUIPT and the density of states in the N region.}} (a) Scheme of the interferometer. $L$ is the length of the proximized normal metal wire (N) while $w$ is the width of the superconducting (S) tunnel junction coupled to the middle of the wire. $\varphi$ is the quantum phase difference in S, whereas $\Phi$ is the externally applied magnetic flux. The tunnel junction normal-state resistance is denoted with $R_t$, $I$ is the current flowing through the device and $V$ is the voltage drop developed across the junction. The spatial coordinate along the N wire is denoted by $x$. (b) Three-dimensional plot showing the evolution of the amplitude (along the vertical axis) of the zero-temperature density of states in the N wire $N_N(\varepsilon,\varphi)$  normalized to the DOS in the absence of proximity effect (i.e., in the normal state) versus energy $\varepsilon$ and phase $\varphi$. The calculation is performed at $x = 0$, i.e., in the middle of the wire. $\Delta_0$ is the zero-temperature superconducting energy gap. 
The latter is a typical feature of a superconductor and indicates the energy interval where there is lack of available states suitable for quasiparticles. 
The coloring of the surface has the role of visually emphasizing the three-dimensionality of the curve. In particular, the blue color indicates where the amplitude of the DOS approaches that in the normal state, whereas the yellow-orange one emphasizes a strong enhancement of the DOS amplitude. 
(c) Zero-temperature $N_N$ versus energy calculated for $\varphi=\pi/2$ at different positions along N. $x$ is the coordinate along the wire, and $x = \pm L/2$ denotes the NS interface boundaries.
}
\label{fig1}
\end{figure}
\section{Proximity effect and device setup}
\label{PE}
Proximity effect is a phenomenon which can be described as the induction of superconducting-like properties into a normal-type conductor thanks to the contact with a superconductor. On the other side, weakening of the correlations typical of a superconductor will occur as a consequence of proximity effect on the superconducting side as well. Yet, as the superconducting state is characterized by a macroscopic quantum phase the latter will affect profoundly such induced correlations. In particular, one relevant consequence of this effect is the modification of the DOS in the normal metal, and the opening of an energy gap (i.e., an energy interval with no energy states available to be occupied by electrons) whose amplitude can be controlled by changing the macroscopic quantum phase of the superconducting order parameter. 
The simplest implementation of a SQUIPT device is shown in Fig. \ref{fig1}(a) and consists of a diffusive (i.e., with dimensions larger than the elastic mean free path) normal metal (N) wire of length $L$ in good electric contact with two superconducting electrodes (S) defining a loop. 
The contact with S therefore induces superconducting correlations in N through proximity effect which is responsible for the modification of the wire DOS.

The following calculations are performed in the short-junction limit where proximity effect in the wire is maximized therefore optimizing the interferometer performance. 
In particular, Fig. \ref{fig1}(b) shows a three-dimensional plot of the amplitude of the N region DOS [$N_N(\varepsilon,\varphi)$, along the vertical axis] calculated in the middle of the wire (i.e., at $x = 0$) as a function of the energy $\varepsilon$ and superconducting phase $\varphi$. 
In particular, $N_N$ is an even function of the energy, and shows an energy gap whose magnitude can be controlled through the quantum phase. 
The energy gap in the DOS turns out to be maximized at $\varphi = 0$, where it obtains its largest amplitude $\Delta_0$, while it is gradually reduced by increasing the phase and is fully suppressed for $\varphi = \pi$. We note that $N_N$ is symmetric with respect to the energy $\varepsilon$, as the DOS physical description is here provided thanks to the aid of the so-called “semiconductor model”. 
In such a model, the density of states of a normal metal is represented as a continuous distribution of single-particle energy levels which includes states both below (i.e., \emph{negative}) and above (i.e., \emph{positive}) the Fermi level, the latter setting the zero of the energy. 
Similarly, a superconductor can be represented with the semiconductor model by symmetrizing its DOS with respect to the energy in such a way that when the energy gap $\Delta_0$ is zero it properly tends to the DOS in the normal state. This representation is particularly useful when computing the tunneling current in system comprising normal metals, superconductor and proximized metals. 
Thus it turns out that a proximized metal behaves as a sort of phase-tunable superconductor, meaning that its energy spectrum can be modified at will with the quantum phase, in particular, by closing or opening the gap in its DOS. 
In addition, the evolution of $N_N$ calculated by changing the position $x$ along the N wire and for $\varphi = \pi/2$ is displayed in Fig. \ref{fig1}(c). 
Furthermore, a superconducting junction (S) of width $w$ and normal-state resistance $R_t$ is coupled through a tunnel junction to the middle of the N wire. 
The loop geometry of the superconducting electrode allows changing the phase difference across the normal metal-superconductor boundaries through the application of an external magnetic field which gives rise to a total flux $\Phi$ through the ring area. 
This modifications of the wire DOS therefore changes electron transport through the tunnel junction. For simplicity we suppose to feed a constant electric current $I$ through the circuit while the voltage drop $V$ developed across the junction is recorded as a function of $\Phi$. 
In the limit that the kinetic inductance of the superconducting loop is negligible, the magnetic flux fixes a phase difference $\varphi = 2\pi \Phi/\Phi_0$ across the normal metal wire, where $\Phi_0  = 2.067\times 10^{-15}$ Wb is the flux quantum.

\begin{figure}[t!]
\includegraphics[width=\columnwidth]{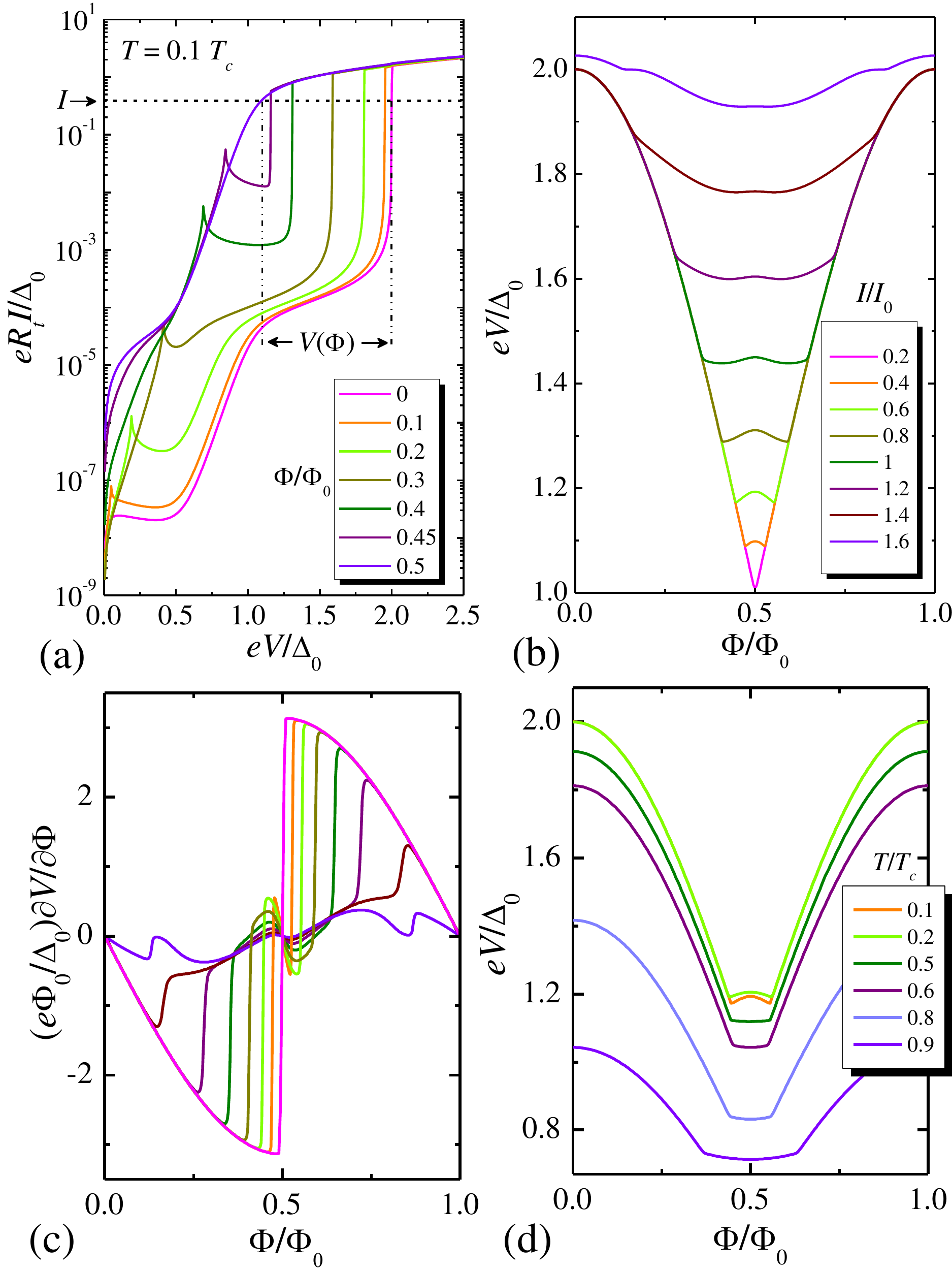}
\caption{\emph{\textbf{SQUIPT predicted behavior.}} (a) Interferometer current vs voltage characteristics ($I-V$) calculated for a few values of $\Phi$ at $T = 0.1T_c$. $\Phi_0$ is the flux quantum, $T_c$ is the superconducting critical temperature, and $e$ is the electron charge. The range of values of $V(\Phi)$ as $\Phi$ is varied over the interval $0 \leq \Phi \leq (1/2)\Phi_0$, for a particular value of the bias current $I$, is also shown. 
(b) Voltage modulation $V(\Phi)$ calculated for several bias currents $I$ at $T = 0.1T_c$. In the figure, $I_0 = \Delta_0/(eR_t)$. (c) Flux-to-voltage transfer function vs $\Phi$ calculated for the same $I$ values and temperature as in panel (b). (d) Voltage modulation $V(\Phi)$ calculated for a few different bath temperatures at $I = 0.6I_0$. In all calculations we set $w = L/3$.
}
\label{fig2}
\end{figure}

\section{Device response}
\label{devres}

Figure \ref{fig2}(a) shows the low-temperature current-voltage ($I-V$) characteristic of the SQUIPT calculated at a few selected values of $\Phi$. It appears that for $\Phi=0$, i.e., when the gap in the N region is fully developed and maximized, the current-voltage characteristic of the interferometer resembles that of a superconductor-insulator-superconductor (SIS) junction composed of two identical superconductors, where the onset of large quasiparticle current occurs for voltages exceeding that corresponding to the sum of the gaps, i.e., for $V\geq 2\Delta_0/e$, where $\Delta_0$ is the zero-temperature superconducting energy gap. The latter is a typical feature of a superconductor, and indicates the energy interval where there is lack of available states for quasiparticles. 
By contrast, for $\Phi=\Phi_0/2$ the gap turns out to be entirely suppressed so that the SQUIPT $I-V$ characteristic corresponds to that of a normal metal-insulator-superconductor (NIS) junction, where the onset for large quasiparticle current occurs for $V\geq \Delta_0/e$. The SQUIPT thus behaves as a \emph{flux-to-voltage transformer} whose response $V(\Phi)$ depends on the bias current $I$ flowing through the tunnel junction.

The interferometer voltage modulation $V(\Phi)$ is shown in Fig. \ref{fig2}(b) for different values of bias current $I$. In particular, $V(\Phi)$ is strongly dependent on the bias current, the latter determining the exact shape of the device response. 
The modulation amplitude of $V(\Phi)$ turns out to be maximized at the lower bias currents where the voltage swing obtains values as large as $\Delta_0/e$, whereas it is gradually reduced by increasing $I$. 
An important figure of merit of the interferometer is represented by the flux-to-voltage transfer function, $F(\Phi)=\partial V/\partial \Phi$, which is shown in Fig. \ref{fig2}(c) for the same $I$ values as in panel (b). In particular, $F(\Phi)$ as large as $\sim3\Delta_0/(e\Phi_0)$ can be obtained at the lowest currents, whereas it is gradually suppressed at higher biasing current.

The role of the temperature on the SQUIPT voltage modulation is shown in Fig. \ref{fig2}(d) which displays $V(\Phi)$ calculated for several temperature values at $I=0.6\Delta_0/(eR_t)$. An increase in temperature leads to a reduction of $V(\Phi)$ as well as to a suppression and smearing of the voltage swing. This reflects analogously on the flux-to-voltage transfer function amplitude.

\begin{figure}[tb]
\includegraphics[width=\columnwidth]{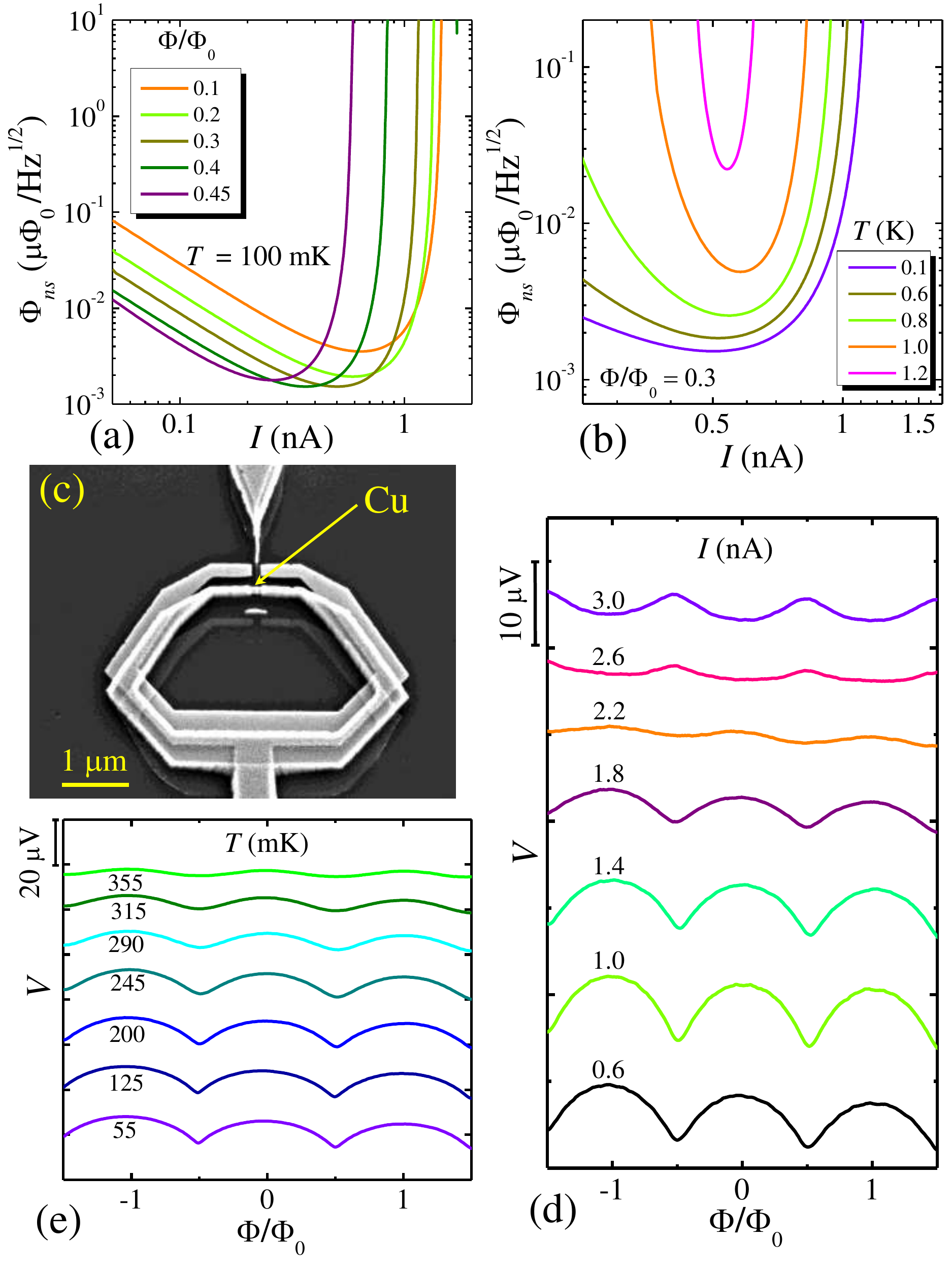}
\caption{\emph{\textbf{Noise performance and behavior of a real SQUIPT device.}} 
(a) Flux resolution $\Phi_{ns}$ vs $I$ calculated for a few $\Phi$ values at $T = 100$ mK. 
(b) Flux resolution $\Phi_{ns}$ vs $I$ calculated at different temperatures for $\Phi = 0.3 \Phi_0$. In these calculations we set $\Delta_0 = 200\,\mu$eV, and $R_t = 200$ k$\Omega$. 
(c) Scanning electron micrograph of a typical Al-based SQUIPT. The ring as well as the probing junction are made of aluminum (Al) whereas the N region is made of copper (Cu). 
The total length of the Cu wire is $L \sim 400$ nm, the ring interelectrode spacing is around $\sim 150$ nm whereas the tunnel probe width is $w \sim 60$ nm. 
(d) Voltage modulation $V(\Phi)$ of a typical SQUIPT measured at $54$ mK  for several values of $I$. The curve are vertically offset for clarity. 
(e) Voltage modulation $V(\Phi)$ measured at a few bath temperatures for $I = 1$ nA. The curves are vertically offset for clarity.
}
\label{fig3}
\end{figure}

\section{Noise performance}
\label{noise}

 We now discuss the noise properties of the SQUIPT. 
 The intrinsic flux noise per unit bandwidth of the interferometer $(\Phi_{ns})$ can be expressed as $\Phi_{ns}=|\partial V/\partial I|S_I^{1/2}/|F(\Phi)|$, where $S_I$ is the current noise spectral density (i.e., the shot noise). 
 In the following we set $\Delta_0=200\,\mu$eV as representative value for a SQUIPT exploiting aluminum (Al) as superconductor, $w=L/3$ and $R_t=200\,\text{k}\Omega$. Figure \ref{fig3}(a) shows $\Phi_{ns}$ versus $I$ calculated for several flux values at $T=100$ mK. 
 In particular, $\Phi_{ns}$ is a non-monotonic function of $I$ with a minimum which depends on the specific value of the applied magnetic flux $\Phi$. 
Moreover, an increase in $\Phi$ leads to a general reduction of $\Phi_{ns}$ at low $I$, while its minimum moves toward lower bias current. 
We stress that $\Phi_{ns}$ down to a few n$\Phi_0/\text{Hz}^{1/2}$ can be achieved, in principle, at low temperature in the $\sim 0.1\ldots 1$ nA range for suitable values of $\Phi$. 
This high flux sensitivity stems from the low shot noise $S_I$ together with a small $|\partial V/\partial I|$ at the biasing point, and large $F(\Phi)$.

The temperature dependence of flux sensitivity is displayed in Fig. \ref{fig3}(b) where $\Phi_{ns}$ vs bias current is plotted for different $T$ values at $\Phi=0.3\Phi_0$. 
Notably, the minimum of $\Phi_{ns}$ turns out to be quite insensitive to the temperature up to $\sim 600$ mK. Then, higher $T$ yields to a reduction of the current window suitable for high flux sensitivity and to an overall enhancement of $\Phi_{ns}$. 
Furthermore, for temperatures larger than 1 K $\Phi_{ns}$ start to be significantly degraded in the whole $I$ range.

Proper operation of the SQUIPT requires the avoidance of magnetic hysteresis in order for the $V(\Phi)$ characteristics to be single valued. This condition can be expressed by stating that $2\pi I_c\mathcal{L}_G <\Phi_0$, where $I_c$ is the critical Josephson supercurrent circulating in the SQUIPT loop, 
and $\mathcal{L}_G$ is the geometric inductance of the ring. 
In particular, for a loop of circular shape and radius $r$ the geometric inductance is proportional to $r$, i.e., $\mathcal{L}_G\sim \mu_0 r$, where $\mu_0$ is the permeability of vacuum. 
From this follows that a small loop radius will allow to suppress its inductance with the consequence of minimizing any possible magnetic hysteresis. 
Therefore a small-diameter SQUIPT could be, in principle, suitable for the investigation of the magnetic properties of small isolated samples. 
As a matter of fact, the magnetometer sensitivity ($S_n$) to an isolated magnetic dipole placed at the center of the loop is approximately given by 
$S_n=2r\Phi_{ns}/(\mu_0 \mu_B)$ where $\mu_B$ is the Bohr magneton. 
It therefore turns out that with a submicron-diameter ring the SQUIPT could provide $S_n\sim 1\ldots 10$ atomic spin/ Hz$^{1/2}$ at temperatures below 1K. 

The technology required for fabricating such nanoscale SQUIPTs is currently widespread, and is based on standard electron-beam lithography (EBL) combined with angle shadow-mask evaporation of metals through a conventional suspended resist mask in a single vacuum cycle. In particular, initially a bilayer of poly(methyl methacrylate)/copolymer resist is spun on an oxidized Si wafer onto which the SQUIPTs devices are patterned using EBL. The structures are then realized by tilting the chip to different angles with respect to the metals sources present in an electron-gun evaporator to deposit the various parts composing the interferometer. The tunnel probe in the device is obtained by exposing the sample to a known pressure of oxygen for a suitable amount of time. This is typically achieved by thermally oxidizing an Al layer which is capable to provide high-quality tunnel junctions.

\section{Real SQUIPT device}
\label{real}

The scanning electron micrograph of a typical SQUIPT device is shown in Fig. \ref{fig3}(c). These structures were fabricated with EBL and angle shadow-mask evaporation of metals, as detailed in the above section. The structure shown in the figure consists of an Al superconducting loop interrupted by a copper (Cu) normal-metal wire. Furthermore, the probing junction appearing in the top of the image, and enabling the device operation, is made of Al as well. The tunneling contact in the present device is made through thermal oxidation of the Al layer of the probe junction leading to a normal-state resistance with typical values of the order of $R_t \sim 50\,\text{k}\Omega \ldots 1\,\text{M}\Omega$.

The $V(\Phi)$ dependence for a typical SQUIPT device with $R_t = 50\,\text{k}\Omega$ measured at 54 mK for several values of the bias current $I$ is shown in Fig. \ref{fig3}(d). As expected [see Fig. \ref{fig2}(b)], the modulation amplitude of $V(\Phi)$  is a non-monotonic function of $I$, while $V(\Phi)$ displays changing of concavity for suitable values of the bias current. 
In this particular sample the voltage modulation obtains values as large as $\sim 7\,\mu$V at 1 nA. 
The corresponding transfer function obtains values as large as $\sim 30\,\mu \text{V}/\Phi_0$ at 1 nA. 
The impact of temperature is shown in Fig. \ref{fig3}(e) which displays the modulation amplitude of $V(\Phi)$ measured at 1 nA in another SQUIPT device for several increasing bath temperature values. In particular, the modulation amplitude of $V(\Phi)$ monotonically decreases by increasing $T$. 

So far SQUIPTs have demonstrated flux-to-voltage transfer function amplitudes up to $\sim 1.5\,\text{mV}/\Phi_0$ leading to flux sensitivities down to $\sim 6\,\mu \Phi_0\text{Hz}^{-1/2}$  below 1 K.  Large improvement of the intrinsic figures of merit of the interferometer is to be expected through a careful optimization of the structure design parameters as well as with suitable cryogenic read-out electronics. Nowadays these interferometers are currently under development from the performance point of view in order to be able to be exploited for the investigation of nanoscale magnetic structures.

Compared to conventional DC SQUIDs, power dissipation ($P$) is dramatically suppressed in the SQUIPT. In these devices $P\sim 100$ fW, which can be further reduced by increasing the resistance of the probing junction. Such a power is $4-5$ orders of magnitude smaller than that in conventional DC SQUIDs, which makes the SQUIPT ideal for applications where very low dissipation is required. We shall finally remark some peculiarities that make this device attractive for a variety of applications: (i) only a simple DC read-out scheme is required, similarly to DC SQUIDs; (ii) either current- or voltage-biased measurement can be conceived depending on the setup requirements; (iii) a large flexibility in the fabrication parameters and materials, such as semiconductors, carbon nanotubes or graphene instead of normal metals, is allowed to optimize the response and the operating temperature; (iv) ultralow dissipation which makes it ideal for nanoscale applications; (v) ease of implementation in a series or parallel array (depending on the biasing mode) for enhanced output.

For background information \emph{see} SUPERCONDUCTIVITY; SEMICONDUCTOR MODEL; JOSEPHSON EFFECT; PROXIMITY EFFECT; MAGNETIC FIELD DETECTION; SUPERCONDUCTING QUANTUM INTERFERENCE DEVICE (SQUID); CRYOGENICS.
\vspace{0.1cm}
\section{Bibliography}

F. Giazotto, J. T. Peltonen, M. Meschke, and J. P. Pekola, Nature Phys. \textbf{6}, 254 (2010).

F. Giazotto and F. Taddei, Phys. Rev. B \textbf{84}, 214502 (2011).

M. Meschke, J. T. Peltonen, J. P. Pekola, and F. Giazotto, Phys. Rev. B \textbf{84}, 214514 (2011).

P. G. de Gennes, \emph{Superconductivity of Metals and Alloys} (W. A. Benjamin, 1966).

M. Tinkham, \emph{Introduction to Superconductivity 2nd edn} (McGraw-Hill, 1996).

J. Clarke and A. I. Braginski (eds), \emph{The SQUID Handbook} (Wiley-VCH, 2004).

\section{More specialized papers}

B. D. Josephson, Phys. Lett. \textbf{1}, 251 (1962).

J. Gallop, Supercond. Sci. Technol. \textbf{16}, 1575 (2003).

W. Belzig, F. K. Wilhelm, C. Bruder, G. Sch\"on, and A. D. Zaikin, Superlatt. Microstruct. \textbf{25}, 1251 (1999).

F. Giazotto, T. T. Heikkil\"a, A. Luukanen, A. M. Savin, and J. P. Pekola, Rev. Mod. Phys. \textbf{78}, 217 (2006).

\end{document}